\documentclass[pdflatex,sn-mathphys-ay]{sn-jnl}

\usepackage{graphicx}%
\usepackage{multirow}%
\usepackage{dcolumn}
\usepackage{amsmath,amssymb,amsfonts}%
\usepackage{amsthm}%
\usepackage{mathrsfs}%
\usepackage[title]{appendix}%
\usepackage{xcolor}%
\usepackage{textcomp}%
\usepackage{manyfoot}%
\usepackage{booktabs}%
\usepackage{algorithm}%
\usepackage{algorithmicx}%
\usepackage{algpseudocode}%
\usepackage{listings}%
\usepackage{amsmath,amssymb,physics,enumerate,graphicx,placeins,quantikz,adjustbox,xcolor}
\usepackage{soul,amsthm,tabularx}

\begin{document}

\title[Article Title]{Detection states of ions in a Paul trap \\ via conventional and quantum machine learning algorithms}

\author*[1]{\fnm{Ilia} \sur{Khomchenko}}\email{ilyakhomka@gmail.com}
\equalcont{These authors contributed equally to this work.}

\author[2]{\fnm{Andrei} \sur{Fionov}}\email{A.Fionov@rqc.ru}
\equalcont{These authors contributed equally to this work.}

\author[3]{\fnm{Artem} \sur{Alekseev}}

\author[3]{\fnm{Daniil} \sur{Volkov}}

\author[1,2]{\fnm{Ilya A.} \sur{Semerikov}}

\author[1,2]{\fnm{Nikolay N.} \sur{Kolachevsky}}

\author[1,2]{\fnm{Aleksey K.} \sur{Fedorov}}

\affil*[1]{\orgname{P.N. Lebedev Physical Institute of the Russian Academy of Sciences}, \orgaddress{\street{53, Leninsky Avenue}, \city{Moscow}, \postcode{119991}, \country{Russia}}}

\affil[2]{\orgname{Russian Quantum Center}, \orgaddress{\street{30, Bolshoy Boulevard}, \city{Moscow}, \postcode{143026}, \country{Russia}}}

\affil[3]{\orgdiv{Center for Artificial Intelligence Technology}, \orgname{Skolkovo Institute of Science and Technology}, \orgaddress{\street{30, Bolshoy Boulevard}, \city{Moscow}, \postcode{121205}, \country{Russia}}}

\abstract{Trapped ions are among the leading platforms for quantum technologies, particularly in the field of quantum computing. 
Detecting states of trapped ions is essential for ensuring high-fidelity readouts of quantum states.
In this work, we develop and benchmark a set of methods for ion quantum state detection using images obtained by a highly sensitive camera.
By transforming the images from the camera and applying conventional and quantum machine learning methods, including convolution, support vector machine (classical and quantum), and quantum annealing, we demonstrate a possibility to detect the positions and quantum states of ytterbium ions in a Paul trap. 
Quantum state detection is performed with an electron shelving technique: depending on the quantum state of the ion its fluorescence under the influence of a 369.5 nm laser beam is either suppressed or not. 
We estimate fidelities for conventional and quantum detection techniques. In particular, conventional algorithms for detecting $^{171}$Yb$^{+}$, such as the support vector machine and photon statistics-based method,as well as our quantum annealing-based approach, have achieved perfect fidelity, which is beneficial compared to standard techniques. This result may pave the way for ultrahigh-fidelity detection of trapped ions via conventional and quantum machine learning techniques.}


\maketitle

\section{Introduction}\label{sec1}

Quantum computing has achieved remarkable progress over the past few decades. 
Recent advances in the development of quantum processors~\cite{debnath2016demonstration, monz2016realization, barends2014superconducting, corcoles2015demonstration} have made it possible to solve certain computationally hard problems with these devices. 
Quantum computational advantage has been demonstrated in several recent experiments~\cite{Arute2019QuantumSU, WuYulin2021, Huang2022, Zhong2020}.

To execute algorithms to solve these challenging tasks, quantum hardware must have high-quality quantum logic gates~\cite{Nielsen_2010book}, as well as readout~\cite{ding2019fast}. 
The trapped ions have become one of the first proposed implementations of qubits~\cite{cirac1995quantum}. 
Their states can be prepared, measured, and entangled efficiently~\cite{monroe2013scaling}. 
It is important that trapped ions can provide high-fidelity (more than $99.93 \%$) and fast (46 $\mu$s per sample~\cite{todaro2021state}) readout~\cite{harty2014high} of quantum states~\cite{kelly2015state}.
Quantum information is encoded in the internal atomic energy levels of the ions, which possess long coherence times, whereas the entanglement between qubits is carried out via the Coulomb force or photonic interconnects~\cite{harty2014high, monroe2013scaling}. 

There are several ion species that are considered to be suitable for quantum computing.  Ions that have a strong optical transition~\cite{monroe2013scaling} reachable for modern laser sources for qubit initialization, laser cooling of the motion, and efficient qubit readout are appropriate. 
Certain simple atomic ions with a single outer electron from alkaline earth metals, such as (Be$^+$~\cite{gaebler2016high}, Mg$^+$~\cite{zalivako2019nonselective}, Ca$^+$~\cite{harty2014high}, Sr$^+$~\cite{reens2022high}, and Ba$^+$~\cite{madej1992observation}) or certain transition metals (Zn$^+$~\cite{yang2020zinc}, Hg$^+$~\cite{hoang2019performance}, Cd$^+$~\cite{brickman2007magneto}, and Yb$^+$~\cite{olmschenk2007manipulation, zalivako2022compact}) satisfy these criteria. Ytterbium ions ($^{171}$Yb$^+$) have become a popular platform for the realization of optical~\cite{Zalivako_2021, zalivako2022compact, Aksenov_2023, zalivako2024towards} and hyperfine qubits~\cite{olmschenk2007manipulation, chen2023benchmarking, moses2023race} as well as optical clocks~\cite{Khabarova_2022, huntemann2012high}.

The detection of the qubit state is usually implemented using an electron shelving method~\cite{leibfried2003quantum}. It relies on a strong dependence of the scattering rate of laser light resonant to some cyclic transition in the ion on the qubit state. Being in one of the qubit states, usually called a bright state, ion efficiently scatters photons, which are registered with a highly sensitive detector. In the other state, called the dark state, the fluorescence is blocked.

An important task is to analyze the raw data collected by a photon detector and restore from it a qubit state, taking into account optical aberrations, background light, detector dark counts and other effects. Unlike traditional methods for qubit-state detection, including the threshold method~\cite{edmunds2021scalable}, the maximum-likelihood method~\cite{langer2006high}, or the adaptive-maximum-likelihood method~\cite{myerson2008high}, machine learning (ML) approaches can guarantee high-accuracy readout in a short detection time~\cite{seif2018machine, ding2019fast}. In this work, we use a set of conventional and quantum ML methods to detect the states of $^{171}$Yb$^{+}$ ions confined in a Paul trap. 
The proposed algorithms distinguish states $\ket{0}=\,^{2}$S$_{1/2}$ $\ket{F=0, m_F=0}$ and $\ket{1}=\,^{2}$D$_{3/2}$ $\ket{F=2, m_F=0}$ of the $^{171}$Yb$^{+}$ ion coupled by a quadrupole transition at 435.5~nm. The readout is performed by illuminating the ion with a laser light at 369.5~nm phase-modulated at 14.7~GHz and a light at 935~nm~\cite{zalivako2024towards}. The ions fluorescence at 369~nm is collected with a high aperture lens on a highly sensitive camera matrix. In this case, the state $\ket{0}$ is bright, rendering the ion appear as a spot on the camera image, while $\ket{1}$ is dark, making the ion invisible in the picture.

A camera produced the dataset of images, each image containing an ion whose state needs to be recognized. Our image quality recognition approach relies on the brightness and positions of the ions derived from input data. Using the performance of conventional and quantum algorithms in terms of fidelity, we provide a highly accurate classification of the ion state. Thus, our work aims to increase the fidelity of the readout in ion-based quantum computers.

This paper is structured as follows. 
In Sec.~\ref{section_2}, we describe the datasets considered in this work. 
Various conventional and quantum ML methods that we analyze and benchmark in this work are presented in In. Sec.~\ref{sec:methods}, which includes the mathematical formulation of these techniques and other technical details.
An alternative method for detecting ion states based on ion image statistics is reported in Sec.~\ref{sec:statistics}. 
This is followed by Sect.~\ref{sec:res_disc}, where we demonstrate the results obtained and highlight interesting observations. 
We summarize our results and conclude in Sec.~\ref{conclusion}.

\section{Dataset}\label{section_2}

\begin{figure}
    \includegraphics[width=0.4\textwidth]{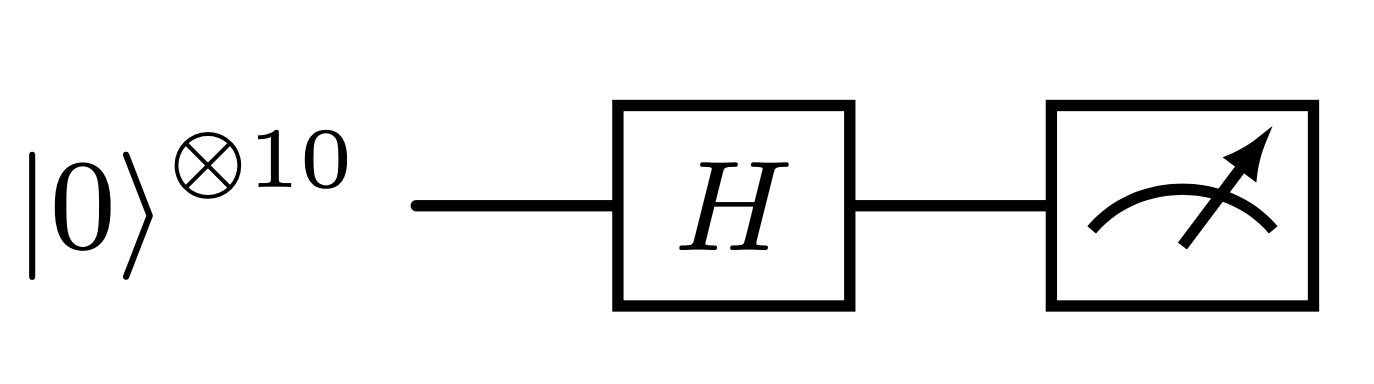}
    \caption{Schematic architecture of the 10-qubit circuit for the preparation of the initial qubits state. }
\label{fig:figure1}
\end{figure}

In order to create a uniform distribution of quantum states, we prepare a 10-qubit circuit to be run on a trapped-ion quantum computer. It consists of Hadamard gates applied to each of the qubits. 
Fig.~\ref{fig:figure1} illustrates the circuit used.

\begin{figure}
    \includegraphics[width=0.4\textwidth]{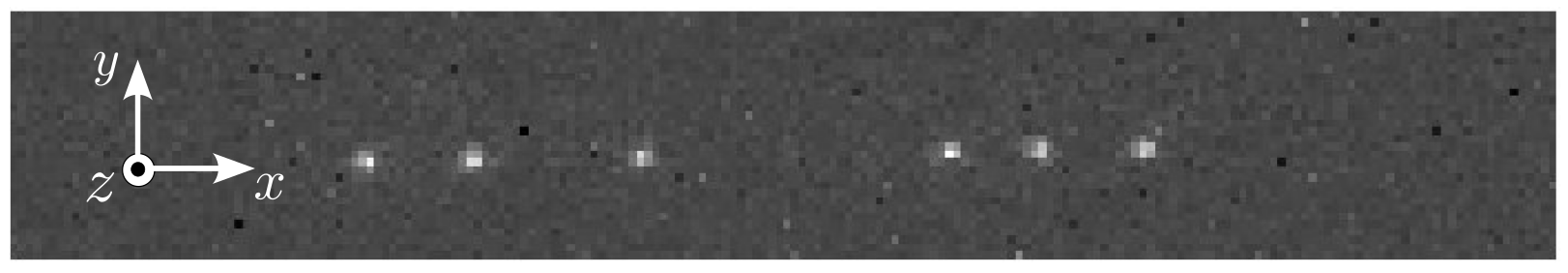}
    \caption{Image of ions, each represented as bright spots, captured with a high-sensitivity camera matrix. The image resolution is 32 × 200 pixels. The chain of ions is aligned along the $x$-axis.}
\label{fig:figure11}
\end{figure}

The first dataset contains 900 images of 10 ions in fluorescent states, known as $Allbright$. 
The second dataset, called the $H_1$ dataset, contains 10,000 images of 10 ions after application of Hadamard operations with an exposition time of 5 ms. Each image in the datasets has a shape of 32 $\cross$ 200 pixels. An example of classified pictures is presented in Fig.~\ref{fig:figure11}.

When ions fluoresce, they become visible in the image obtained with the camera, appearing as bright spots. Following this, we label the dataset $H_1$ such that bright images correspond to ions in the sate $\ket{0}$, while dark images correspond to the state $\ket{1}$. 

\section{Ion Image Statistics}
\label{sec:statistics}
\begin{figure}[ht]  
    \includegraphics[width=0.5\textwidth]{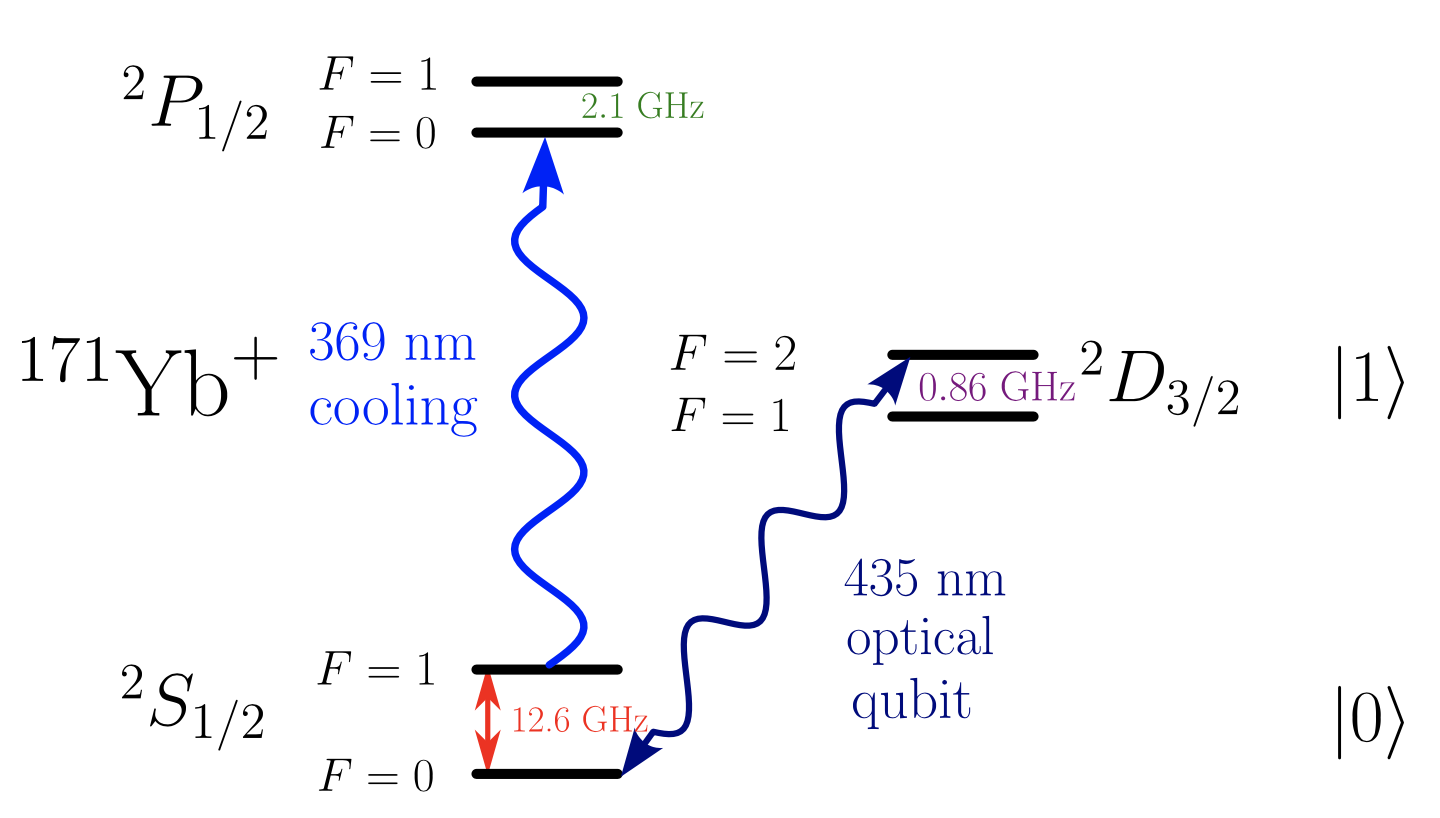}
    \caption{Scheme for detecting the states of the ion $^{171}$Yb$^{+}$. The qubits are encoded in the states $\ket{0}=\,^{2}$S$_{1/2} (F=0, m_F=0)$ and $\ket{1}=\,^{2}$D$_{3/2} (F=2,m_F=0)$, controlled by a laser at 435.5 nm. Cooling, initialization and readout are performed on a transition at 369.5 nm. Depending on the state of the qubit, ion scattering at 369.5 nm is either allowed or blocked, and this is detected with a camera.}
    \label{fig:figure5}
\end{figure}

Each image in the datasets contains a chain of $^{171}$Yb$^{+}$ ions confined in a linear Paul trap. The qubits are encoded in states $\ket{0}=\,^{2}$S$_{1/2} (F=0, m_F=0)$ and $\ket{1}=\,^{2}$D$_{3/2} (F=2,m_F=0)$ and are controlled using a laser beam at 435.5 nm. For a qubit, the state $\ket{0}$ is referred to as the bright state, which corresponds to a strong fluorescence signal under illumination with a 369.5-nm laser beam, which makes the ion visible on the camera. The state $\ket{1}$ is called the dark state, as fluorescence during the readout process is suppressed, rendering the ion invisible in the camera image. Following Doppler cooling performed using a laser beam at 369.5 nm along with a repumping beam at 935 nm, appropriately phase-modulated~\cite{zalivako2024towards}, the ion is initialized into the qubit state $\ket{0}$. Figure~\ref{fig:figure5} shows a schematic representation of the detection process for the qubit states of $^{171}$Yb$^{+}$.  

\begin{figure}[ht]
    \centering    
\includegraphics[width=0.45\textwidth]{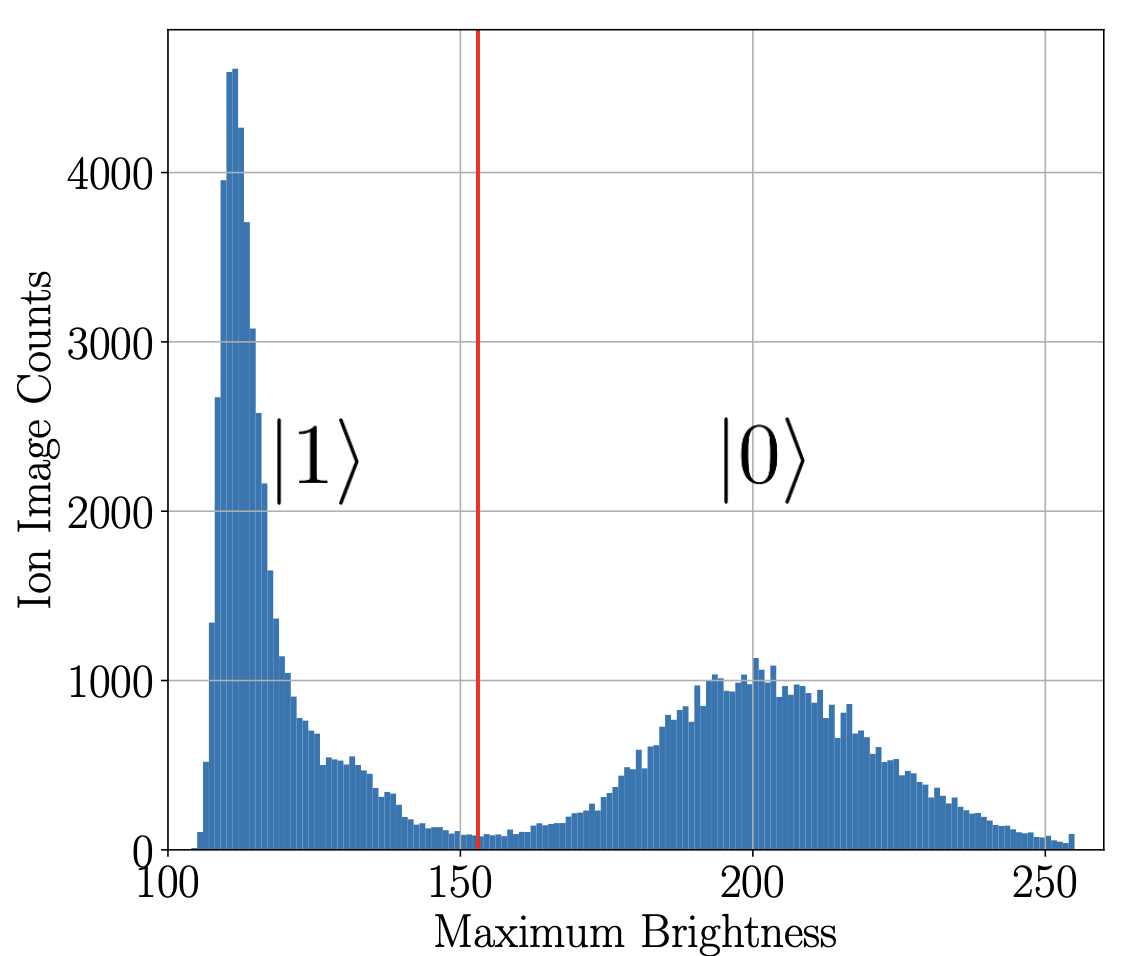}
    \caption{Distribution of the maximum brightness in an anchor box. The red line represents the threshold used to distinguish ions in the two classes, $\ket{0}$ and $\ket{1}$. The maximum brightness of pixels varies from 0 to 255.}
    \label{fig:figure6}
\end{figure}

State-dependent fluorescence detection is used to realize the qubit readout. The histogram of registered ions with the brightest pixels, corresponding to "bright" ions, is plotted as a function of the maximum brightness for a single anchor box from an image in the dataset $H_1$ and shown in Figure~\ref{fig:figure6}.

The statistical algorithm is based on the threshold approach for counting photon transitions to determine the positions of ions within regions of interest, called anchor boxes with images of ions. From a physics perspective, the number of photons can be simulated using their statistics~\cite{semenin2021optimization}. The red line indicates the threshold, set to 153, which separates ions into two classes: $\ket{0}$ and $\ket{1}$.

\section{Methods} 
\label{sec:methods}

Humans often struggle to accurately classify time-averaged images, whereas computation methods may excel. Conventional and quantum ML methods can perform the classification of ions in a Paul trap. To implement the ion state readout, we apply the following procedure. Initially, ions are divided into 10 clusters for the $Allbright$ dataset to detect the center of ions on the images. This is followed by the ion state detection for the other dataset, $H_1$, by means of several algorithms.
For each of the datasets, we automatically labeled the data for the ion states readout with the proposed approaches. This was done by applying the ion-image statistics algorithm, selecting the ion with the maximum brightness in each image, separated by a threshold value, (See Section \ref{sec:methods}) to generate a dataset of labels, with respect to which we assess  the prediction quality of our methods. The readout process concludes with fidelity assessment the between labeled data and predicted results obtained from the $H_1$ dataset.  

The classification algorithm is based on the equilibrium positions of the ions in a trap~\cite{james1998quantum}. 
In this model, a linear chain of $N$ ($N=10$) ions is assumed to be strongly confined in the $y$ and $z$ directions to neglect the motion of the ions in these directions; 
while weakly trapped by a harmonic potential in the $x$ direction. Counting the ions from left to right, we denote the position of the  $m$-th ion as $x_m(t)$.  
The moving ions are influenced by the harmonic trap and by mutual Coulomb interactions, as each ion is located within the electric field of the others. The potential energy of such an ion chain is given by:
\begin{equation}
   V = \sum_{m = 1}^{N}\frac{1}{2}M^{2} \nu^2 x_m (t)^{2} + \sum_{n, m = 1; n \neq m}^{N}\frac{Z^{2} e^{2}}{8 \pi \eta_{0}}\frac{1}{|x_n(t)-x_m(t)|},  
\label{eq:pe}
\end{equation}
where $M$ denotes the mass of each ion in the chain, 
$e$ is the electron charge, 
$Z$ is the degree of ionization of the ions, 
$\eta_{0}$ is the permittivity of free space, and $\nu_0$ is the trap frequency, which characterizes the strength of the trapping potential in the axial direction. For sufficiently cold ions, their positions are determined as follows (see Ref.~\cite{james1998quantum}):
\begin{equation}
    x_m(t) = x_m(0) + q_m(t), 
\label{eq:ip}
\end{equation}
where $x_m(0)$ and $q_m(t)$ denote the equilibrium position of the ion and a small displacement, respectively. 
The equilibrium positions of the ion are obtained according to the following criterion
\begin{equation} 
    \left[\frac{dV}{dx_m} \right]_{x_m=x_{m}(0)}=0.
\label{eq:eq}
\end{equation}

\begin{figure}[h] 
    \includegraphics[width=0.45\textwidth]{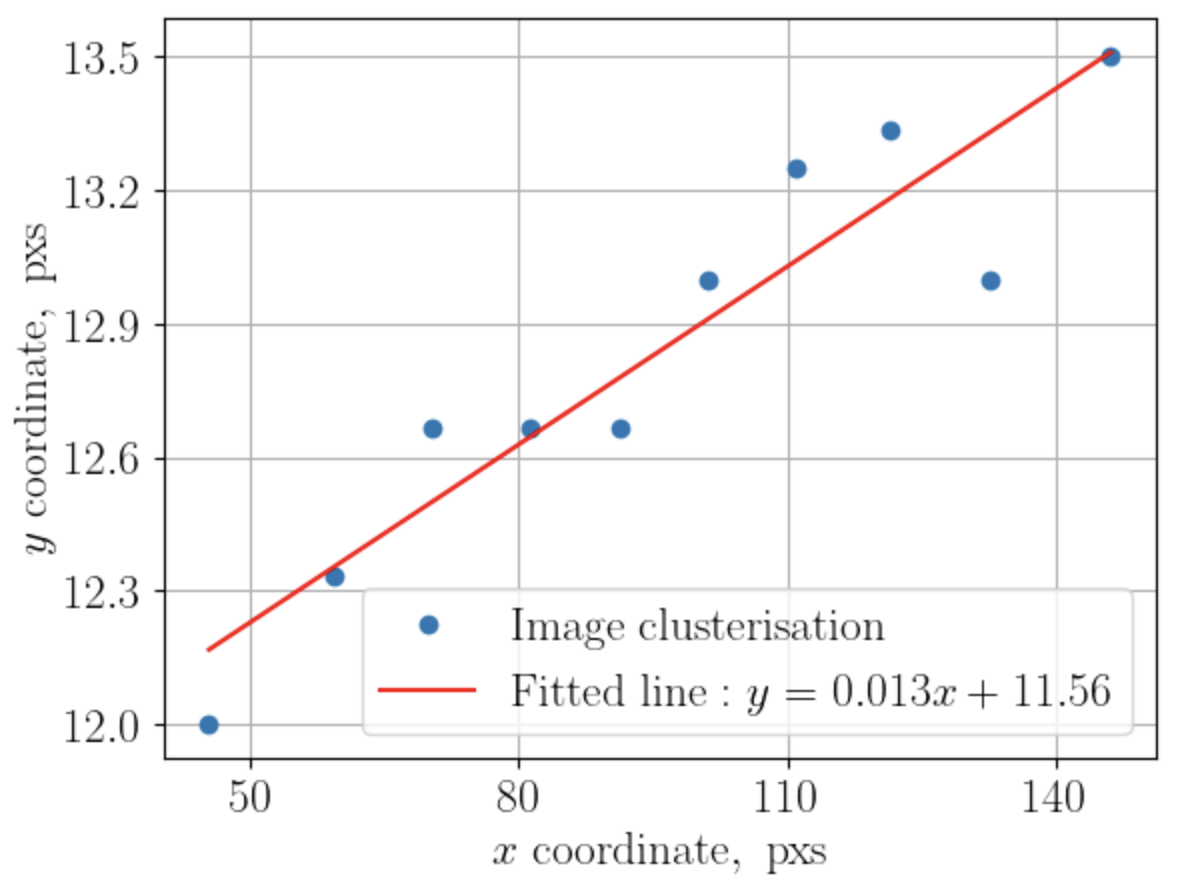}
    \caption[Coordinates of ions]{Coordinates of ions determined using the k-means method (blue dots) and the fitting line (red), as the ions are located along a straight line.  In this case, we consider $k=10$ clusters to establish the coordinates of ions, while $k=2$-clustering is applied to distinguish between bright and dark states. The coordinates of the ions are measured in pixels.}
    \label{fig:figure3}
\end{figure}

The ion positions in 900 images were determined using the $k$-means algorithm from the Python library scikit-learn~\cite{pedregosa2011scikit}. 
Applying Eq.~\eqref{eq:ip} and the $k$-means method, we derived the coordinates of the ions for the images in the $Allbright$ dataset, which were fitted by the regression line $y = m x + b$, as shown in Fig.~\ref{fig:figure3}.

The coordinates obtained are then passed to the next steps of the algorithm, where they are flattened from a matrix into a feature vector based on pixel brightness. 
The feature set includes the following parameters: maximum brightness, minimum brightness, the difference between the maximum and minimum brightness values, mean brightness, median brightness, standard deviation, variance, skewness, kurtosis, fundamental tone frequency, and maximum power for the fundamental frequency\footnote{The fundamental frequency is the lowest frequency of a periodic function.} derived from the Fourier transform. 

The first used approach is the supervised binary classification of individual ion images with two classes $-$ bright and dark. The second approach involves unsupervised binary clustering of these images with the same ``bright'' and ``dark'' classes. One can also perform binary classification for such data accounting for the statistical and physical properties of ions. While we automatically labeled the data for classification (so-called supervised learning), labels for a test dataset were included only to avoid associating an ion with the wrong class. For algorithms, such as $k$-means, which rely on the physical characteristics of ions or the distribution of bright pixels, automatic labeling is necessary. To demonstrate the advantages of quantum algorithms, we investigated them alongside their conventional counterparts.

\subsection{Supervised Learning}

\subsubsection{Support vector machine}

A support vector machine (SVM) is used for a variety of tasks, including image classification, because this method achieves high accuracy even for a small size of a dataset~\cite{Foody2004ARE, Mathur2008MulticlassAB}.
Let us consider a training dataset that contains $l$ points $\{x_i, y_i\}$, where $x_i$, $i=1, \dots, l$ represent the features of the ions' anchor boxes, $y_i \in \{-1, 1\}$ indicates the class to which the point $x_i \in \mathbf{R}^d$ belongs. Here, $d$ denotes the dimensionality of the feature space. In the most general case, when the data can be separated by only a nonlinear decision surface, we can map these data to some other high-dimensional Hilbert space $\mathcal{H}$ as $\Phi: \mathbf{R}^d \xrightarrow{} \mathcal{H}$. To reduce the expensive computation of $(\Phi(x) \cdot \Phi(x_i))$, we apply a positive definite kernel introduced as 
\begin{equation}
K(x, x_i) =(\Phi(x) \cdot \Phi(x_i)). 
\label{eq:kernel}
\end{equation}

This classifier aims to fit an optimal separating hypersurface between classes. Specifically, it seeks a hypersurface that separates vectors $x_i$ from different classes without errors and minimizes the norm of the coefficients $\alpha_i \geq 0$. To achieve this, one must solve the following optimization problem~\cite{vapnik2013nature}:
\begin{equation}
\max_{\alpha} W(\alpha) = \max_{\alpha_i} \left (  \sum_{i}^{l} \alpha_{i} - \sum_{i}^{l} \alpha_{i} \alpha_{j} y_{i} y_{j}  K(x_{i},x) \right),
\label{eq:ohs}
\end{equation}
subject to the constraints
\begin{equation}
\sum_{i}^{l} \alpha_{i} y_{i} = 0, i =1, \ldots, l. 
\label{eq:con}
\end{equation}
The binary classes $\{-1\}$, $\{1\}$ are then converted to the ion states $\ket{0}$ and $\ket{1}$, respectively. 

A quantum version of the support vector machine, known as a quantum support vector machine (QSVM), performs the same task as its conventional counterpart, with the key that a classical kernel, $K(x_{i},x)$, is replaced by a quantum one $K(x_{i},x) = \braket{\Phi(x)}{\Phi(x_i)}$, where $\ket{\Phi(x)}$ represents the wave function of the quantum system~\cite{suzuki2024quantum}. Our work presents an implementation of the quantum support vector machine based on quantum annealing. This approach includes a binary representation of the model parameters~\cite{Willsch2019} to encode quantum states.

\subsubsection{Convolution algorithm}

The classification of ions in a Paul trap can be solved by purely conventional methods. One of the simplest approaches for this task is convolution. Let $\{x_i, y_i\}$ represent a dataset, where $x_i$, $i=1, \ldots, l$ are the features of the anchor boxes of the ions, and $y_i \in \{0, 1\}$ indicates the class to which each point $x_i \in \mathbf{R}^d$ belongs. If the dataset is labeled using the functionality of the Python library \textit{scikit-learn}~\cite{pedregosa2011scikit}, this procedure enables the computation of a convolution for each ion box. Specifically, by selecting a kernel (reference vector), as the first element of the dataset $x_0$, the convolution with a target vector $x_i$ then reads
\begin{equation} 
\widetilde{y}_i =  \left [1 - \frac{\sum_{i=1}^{l}x_{0,n} x_{i,n-m}}{||x_0|| ||x_i||} \right ],  
\label{eq:conv}
\end{equation}
where $||\cdot||$ denotes the Euclidean norm and $\widetilde{y}_i\in \{0, 1\}$ is a binary variable that indicates the class to which the element $x_i$ belongs. If $\widetilde{y}_i \geq\ \theta$, then $y_i$ = 1, otherwise $y_i$ = 0, where $\theta$ represents the threshold.
There is only one hyperparameter $-$ the threshold, $\theta$, which can be optimized and in the optimal case it is equal to $\theta =0.001523$. As previously, we convert the binary states to the states of the ion as follows: $y_i = 0$ corresponds to $\ket{1}$, and $y_i = 1$ corresponds to $\ket{0}$. 

\subsection{Unsupervised Learning}
\subsubsection{k-means}

$k$-means clustering belongs to the class of unsupervised learning algorithms~\cite{sinaga2020unsupervised}. As a distance-based method, $k$-means clustering groups $l$ objects into $k$ clusters by minimizing the within-cluster sum of squares, defined as: 
\begin{equation}
\sum_{i=1}^{l}\sum_{j=1}^{k} w_{ij} ||x_i-c_j||^2,  
\label{eq:ksum}
\end{equation}
where $\{x_i\}$ is a dataset, $i=1, \ldots, l$, $c_j$ represents a centroid for cluster $j$, ($j=1, \ldots, n$), $w_{ij}$ is a binary variable that equals 1 if point $i$ belongs to cluster $j$ and 0 otherwise, and $||\cdot||$ denotes the Euclidean norm~\cite{burges1998tutorial}. Within each of the $k$ clusters, the objects are more similar to each other than to objects in other clusters.

The algorithm for determining the final set of clusters can be divided into the following steps. First, we choose the number of clusters $k$, which is set to 2 in our case. Then, $k$ random points are selected as initial centroids before partitioning each data point into the closest cluster based on the distance of the data point to the centroid $-$ the Euclidean distance. Next, we calculate the new centroids by averaging the coordinates of all points within each cluster. Lastly, we repeat the previous two steps until the cluster assignments no longer change (no transitions between clusters). As mentioned above, there are two classes: the bright and dark states of the ions, denoted by $\ket{0}$ and $\ket{1}$, respectively, and a list of features, such as the brightness and positions of ions, upon which the classification process is based.

\subsection{Quantum Annealing}
\label{sec:quantum}
Quantum annealing can serve as a tool to classify ions images. The idea of this method relies on adiabatic quantum computing, where the solution to a given problem is encoded in the ground state of a Hamiltonian, $H_{\rm 0}$, which governs the dynamics of a system of $N$ qubits~\cite{schuld2021machine}. If a quantum system evolves adiabatically from the known ground state of $H_{\rm 0}$ to the unknown ground state of $H_{\rm P}$, for sufficiently large $T$, the solution to this computational problem will tend to the ground state of $H_{\rm P}$~\cite{farhi2000quantum}. This evolution at time $t$ is usually approximated by the following Hamiltonian~\cite{lucas2014ising}
\begin{equation}
H(t) = (1 - \frac{t}{T}) H_{\rm 0} + \frac{t}{T} H_{\rm P},  
\label{eq:ham}
\end{equation}
where $H(t=0)= H_{\rm 0}$ and $H(t=T)= H_{\rm P}$.

To implement a device that uses quantum annealing to solve hard problems, one could use a programmable quantum spin system, where each spin and its couplings could be controlled~\cite{johnson2011quantum}. An Ising model Hamiltonian, $H_{\rm P}$, can describe the behavior of this system and is given by 
\begin{equation}
H_{\rm P} = -\sum_{i=1}^{N} h_i \sigma^z_i -\sum_{1 \leq i \leq j}^{N} J_{ij} \sigma^z_i \sigma^z_j,  
\label{eq:Ising}
\end{equation}
where $\sigma^z_i$ is the Pauli spin matrix for spin $i$, $h_i$ is the local magnetic field, and $J_{ij}$ is the coupling energy between the spins $i$ and $j$. The operators $\sigma^z_i$ assign values $\{ \pm 1\}$ to spin states $\{ \uparrow, \downarrow \}$. At the beginning of a computation, the Hamiltonian, $H_{\rm 0}$, corresponds to a transverse magnetic field~\cite{boixo2013experimental}
\begin{equation}
      H_0 = - \sum_{i=1}^{N} \sigma_{i}^{x}, 
\label{eq:trans}
\end{equation} 
with $\sigma^x_i$ being the $x$ Pauli matrix. 

\subsection{Quadratic Unconstrained Binary Optimization}
\label{sec:qubo}
In practice, one can map the Ising Hamiltonian to a Quadratic Unconstrained Binary Optimization (QUBO) representation by replacing $\sigma_{i}^{z}$ by $2x_i - 1$, where $x_i \in \{0, 1\}$, which results in the following form for $H_{\rm P}$
\begin{equation}
H_{\rm P}^{\rm QUBO} = \sum_{i=1}^{N} \sum_{j=1}^{i} Q_{ij}x_i x_j + C, 
\label{eq:QUBO}
\end{equation}
where $Q$ is the QUBO matrix given by
\begin{align}
    Q_{ij} &= \begin{cases}
      -4J_{ij} &\text{if} \ i=j \\
      \sum_{i} \sum_{1\leq k \leq j} (2J_{ik}+2J_{ki}+2h_i) &\text{if} \ i \neq j
    \end{cases}
\end{align}
and $C = \sum_{i} \sum_{1\leq i \leq j}J_{ij} - \sum h_i$ is a constant. The QUBO optimization problem consists of the maximization of $H_{\rm P}^{\rm QUBO}$ with respect to $x$, namely, 
\begin{equation}
      \max_{x} H_{\rm P}^{\rm QUBO}(x) = \max_{x_i} \sum_{i=1}^{N} \sum_{j=1}^{i} Q_{ij}x_i x_j + C, 
\label{eq:qubo}
\end{equation}
subject to the constraints $x_i \in \{ 0, 1\}$.

\subsubsection{``Quant'' algorithm}

Quantum algorithms can be used to classify images using their parameters. Following this, we develop a method, called \textit{``Quant"} for brevity, to illustrate this point. For an image of each ion in the chain, the maximum brightness, $\sigma_i$ $(i=1,2, \dots, N)$, is computed. This process is followed by the calculation of the QUBO matrix for every $\sigma_i$, namely
\begin{equation}
Q_i
=
\left(
\begin{array}{cc}
\sigma_i & \frac{\sigma_i+\varepsilon_i}{2}\\
\frac{\sigma_i+\varepsilon_i}{2} & -\varepsilon_i
\end{array}
\right),
\label{eq:qi}
\end{equation}
where $\varepsilon_i$ is the threshold. The matrix obtained $Q_i$ is then supplied to a solver, which results in the state vector of the chosen ion. Since the $N$ ions on the images are independent, their final states become $\ket{x}_1\otimes  \ket{x}_2 \otimes \ldots  \otimes \ket{x}_N =\ket{x_1 \ldots x_N}$, that is, either $0$ or $1$. For each dataset, we then automatically label the data for the ion states readout with the approach based on ion image statistics.

Interestingly, a matrix of $Q_i$ can be decomposed into a combination of unit gates. Indeed, we obtain 
\begin{equation}
Q_i
= \frac{\sigma_i-\varepsilon_i}{2} (I - \sigma^z) + \frac{\sigma_i+\varepsilon_i}{2} \sigma^x, 
\label{eq:deq}
\end{equation}
where $I$ denotes the identity matrix.

\section{Results and Discussion}
\label{sec:res_disc}

For each of the algorithms presented, we access their performance using fidelity, $\mathcal{F}$, as the figure of merit. Since various authors define fidelity using different approaches, we provide its expression to avoid confusion. Let $\rho_0$ and $\rho_1$ be the initial and final quantum states of a quantum system on a finite-dimensional Hilbert space. Then, the fidelity between these states reads~\cite{Nielsen_2010book}
\begin{equation}
\mathcal{F} =\mathrm{Tr} \ \sqrt{\rho_0^{1/2}\rho_1\rho_0^{1/2}},
\label{eq:fid_quant}
\end{equation}
where $\mathrm{Tr} \{ \cdot \}$ is the conventional trace operator. In the case of classical systems, the fidelity between two probability distributions $\{ p_i \}$ and $\{ q_i \}$ ($i=1, \dots, N$) corresponding to two random variables $X$ and $Y$, respectively, which can take values in $(1, \dots, N)$, is given by~\cite{Nielsen_2010book}
\begin{equation}
\mathcal{F} = \sum_{i=1}^{N}\sqrt{p_i q_i},
\label{eq:fid_class}
\end{equation}
where we follow the classical fidelity definition consistent with the quantum one defined by Eq.~\eqref{eq:fid_quant}~\cite{Nielsen_2010book}. In our case, $\{ p_i \}$ is the probability distribution calculated by the algorithms, and $\{ q_i \}$ is the distribution of labels.

This quantity is a convenient measure for characterizing the performance of conventional and quantum algorithms because fidelity depends only on the initial and final states of the system.

Some of the presented conventional and quantum algorithms require the optimization of hyperparameters to achieve better performance. For example, we optimized the value of the maximum brightness for the algorithm based on ion image statistics, setting it to 153. The optimization procedure for the convolution-based algorithm resulted in $\theta = 0.012$. Using a similar technique, we obtained $\varepsilon_i = \varepsilon = 152.8$ for the algorithm based on image parameters, which we later refer to as \textit{``Quant''}. For all these methods, the $f_1$ score was selected as the cost function for optimization.

Table~\ref{tab:res} presents a summary of our results. All the calculations were performed on an Intel i5-12500H processor with a 2.5 GHz CPU.

\begin{table}
	\caption{Performance of conventional and quantum algorithms. Bold text indicates the best score.}
	\centering
	\begin{tabular}{lll}
		\toprule
		\multicolumn{2}{c}{Conventional algorithms}                   \\
		\cmidrule(r){1-3}
		Algorithm     & Fidelity (\%)  & Time (s) \\
		\midrule
		Convolution  & 99.20 &  0.72  \\
		Support Vector Machine & \textbf{100.00} & 1.93 \\
		$k$-means Clustering       & 99.76 & 0.035  \\
        Ion Image statistics       & \textbf{100.00} & 0.094 \\
            \bottomrule
            \multicolumn{2}{c}{Quantum algorithms}                   \\
            \midrule
		Quant algorithm  & \textbf{100.00} & 77.99   \\
		Quantum Support Vector Machine & 99.41 & 2448.06      \\
            \bottomrule
	\end{tabular}
	\label{tab:res}
\end{table}

Remarkably, the SVM demonstrated the highest fidelity among all the proposed methods, achieving $100 \%$, allowing us to perfectly distinguish the states of ions, but its computational time was the longest among conventional ML methods at 1.93 s. $k$-means clustering exhibited the fastest computational time, with a value of $35$ ms. Interestingly, the approach that relies on ion image statistics outperformed other methods, i.e., it has the second-fastest computational time (94 ms) and the maximum fidelity $100.00 \%$, obtained using the maximum value of brightness as a feature.\footnote{By performance of an algorithm, we mean a trade-off between its computational time and fidelity. The algorithm with the best performance is one whose fidelity is high compared to other methods and, at the same time, has a relatively short computational time} Thus, instead of performing complicated data processing and then providing it to ML approaches, a simple method based on ion image statistics can serve as an attractive option for ion classification. 

Although quantum algorithms concede their conventional counterparts in computational time, the fidelity of the so-called \textit{ ``Quant''} algorithm is $100.00 \%$, which is the highest score among all considered techniques, demonstrated alongside the conventional SVM and ion-image-statistics-based approach. This suggests that our algorithm can be used as an alternative to conventional ML methods to detect ion states within a reasonable time frame. Despite the high fidelity of QSVM ($\mathcal{F} = 99.41\%$), the time to run this algorithm far exceeds the computational times of the other methods due to the substantial time required to compute the quantum kernel upon which QSVM is based.
The implementation of the so-called \textit{``Quant''}algorithm was carried out using the simulated quantum annealing libraries ``SimCIM'' and ``pyqiopt''~\cite{Tiunov2019}, while to realize QSVM, we formulated an SVM problem in the language of quantum annealing as in reference~\cite{Willsch2019} and then applied a D-Wave quantum annealer~\cite{Willsch2019}.  

Absolute recognition of the ion states of $^{171}$ Yb$^{+}$ ($\mathcal{F} = 100.00\%$) using QSVM is feasible. To achieve this result, the size of the training dataset should be increased to $10 \%$ of the entire dataset selected for the conventional SVM in this work. However, due to the large size of the quantum kernel (25000 $\cross$ 25000) and our limited computational resources~\footnote{The computational time depends on binarization parameters and the size of a training dataset.}, we opted for a training dataset size of $5 \%$ for the QSVM. For such a dataset, the size of the QUBO matrix is 25000 $\cross$ 25000, which is already large for simulated quantum annealing. Interestingly, $100 \%$ recognition of images using a quantum-enhanced Support Vector Machine algorithm has recently been reported for a dataset of 10 images in~\cite{zalivako2024supervised}, making quantum versions of the SVM promising candidates for ion state detection.

Previously, convolution neural networks (CNNs) have been recommended as promising candidates to implement a fast and high-fidelity readout of ion states (accuracy $\approx 99.4 \% $)~\cite{Ding_2019}, while, in our case, a simple convolution was applied for this purpose, allowing us to efficiently complete this process, with the fidelity $\mathcal{F}=99.2 \%$, where an ion with the highest brightness was selected as a feature of this mathematical method. Since basic conventional methods determine the position of ions almost without errors, this result omits applying more advanced techniques, such as neural networks. 

\section{Concluding remarks} 
\label{conclusion}

In this paper, we have presented certain conventional and quantum methods to determine the state of ions in a Paul trap. Importantly, the best score (fidelity) was achieved by the SVM algorithm and the ion-image-statistics-based approach (100.00 \%) among conventional algorithms and by our \textit{``Quant''} algorithm (100.00 \%) among quantum approaches, allowing perfect recognition of the states of ions. 
The latter algorithm can be easily scaled as it is based on a $2 \cross 2$ QUBO matrix, which can be presented as a decomposition of single-qubit gates. 
Another significant outcome of this work is that the maximum brightness showed high importance across all algorithms, including those based on ion image statistics and machine learning. Quantum algorithms achieved fidelity nearly equivalent to their conventional counterparts, as both rely on the same algorithmic kernels; however, the determination of ion states takes longer in the quantum case. Our result may improve the fidelity of the readout procedure in quantum computers based on atomic ions. A better understanding of the physics of the photon transition between the $^{2}$S$_{1/2}$ $\ket{F=0}$ and $^{2}$D$_{3/2}$ $\ket{F=2}$ levels may further enhance detection algorithms. Accounting for quantum features in images when applying quantum algorithms might also impact the accuracy and runtime of these methods.

\backmatter

\bmhead{Acknowledgements} We are deeply indebted to the Lebedev Physical Institute of the Russian Academy of Sciences for their experimental support of our work, as well as to Dr. I. Zalivako and Dr. E. Kiktenko for his critical assessment of our manuscript.

\bmhead{Author contribution} I. K. conceptualized the manuscript, designed the methodology, and performed the formal analysis. A. F. participated in the conceptualization and analysis, and wrote the code. A. A. and D. V. assisted with the initial analysis, contributed to the conceptualization of the problem, and helped propose and test algorithms for the initial problem. I. A. S. and N. N. K. provided financial and technical resources for the project, and offered feedback on the first draft of the manuscript. A. K. F. reviewed the initial draft of the paper and supervised the entire project.

\bmhead{Data availability} The data is available in the \hyperref[GitLab]{https://gitlab.com/fionandrey/ion-state-readout.git} repository.

\bmhead{Code availability} The code is available in \hyperref[GitLab]{https://gitlab.com/fionandrey/ion-state-readout.git} repository.

\section*{Declarations}

\bmhead{Conflict of interest} The authors declare no competing interests.

\bibliography{ions}

\end{document}